\documentclass[conference]{IEEEtran}
\IEEEoverridecommandlockouts
\usepackage{cite}
\usepackage{amsmath,amssymb,amsfonts}
\usepackage{algorithmic}
\usepackage{graphicx}
\usepackage{textcomp}
\usepackage{xcolor}
\def\BibTeX{{\rm B\kern-.05em{\sc i\kern-.025em b}\kern-.08em
    T\kern-.1667em\lower.7ex\hbox{E}\kern-.125emX}}
\begin{document}

\title{Codec Data Augmentation for Time-domain Heart Sound Classification\\
% \thanks{This work was supported by }
}

\author{
\IEEEauthorblockN{Ansh Mishra}
\IEEEauthorblockA{\textit{Department of Physics} \\
\textit{IIT Delhi}\\
New Delhi, India \\
ph1210821@iitd.ac.in}
\and
\IEEEauthorblockN{Jia Qi Yip}
\IEEEauthorblockA{\textit{Interdisciplinary Graduate Programme} \\
\textit{Nanyang Technological University (NTU)}\\
Singapore \\
}
\and
\IEEEauthorblockN{Eng Siong Chng}
\IEEEauthorblockA{\textit{School of Computer Science and Engineering} \\
\textit{Nanyang Technological University (NTU)}\\
Singapore \\
}
}

\maketitle

\begin{abstract}
Heart auscultations are a low-cost and effective way of detecting valvular heart diseases early, which can save lives. Nevertheless, it has been difficult to scale this screening method since the effectiveness of auscultations is dependent on the skill of doctors. As such, there has been increasing research interest in the automatic classification of heart sounds using deep learning algorithms. However, it is currently difficult to develop good heart sound classification models due to the limited data available for training. In this work, we propose a simple time domain approach, to the heart sound classification problem with a base classification error rate of 0.8 and show that augmentation of the data through codec simulation can improve the classification error rate to 0.2. With data augmentation, our approach outperforms the existing time-domain CNN-BiLSTM baseline model. Critically, our experiments show that codec data augmentation is effective in getting around the data limitation. 
\end{abstract}


\section{Introduction}
Cardiovascular diseases are a leading cause of death around the world~\cite{GBD2022}. Out of the many cardiovascular diseases, valvular heart disease is a common type of life-threatening disease~\cite{morsi2012tissue} and early detection plays a key role in improving patient outcomes~\cite{wang2015early}. Many cardiac conditions, especially valvular heart diseases, are first picked up on cardiac auscultation. The purpose of cardiac auscultation is to characterize heart sounds and murmurs which can indicate CVDs. With the rise of digital stethoscopes that can convert heart sounds into digital phonocardiogram (PCG) signals for storage and analysis, there have also been efforts to perform automated classification of heart sounds. Compared to other techniques for detecting heart murmurs such as Echocardiogram, Cardiac magnetic resonance imaging, and computed tomography scans, collecting a PCG signal through a digital stethoscope has significant cost advantages~\cite{leng2015electronic}. As such, PCG-based classification of heart sounds remains an important avenue of research.
\begin{figure}[ht!]
  \centering
  \includegraphics[width=180pt]{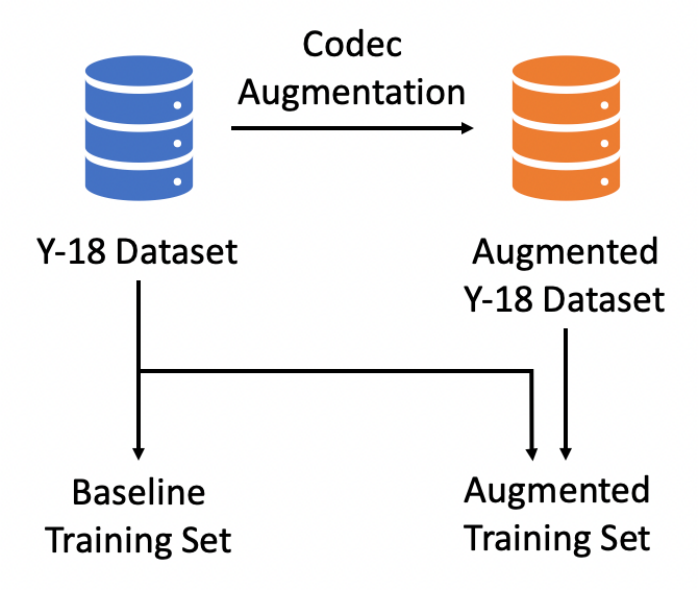}
  \caption{The codec data augmentation strategy. The original data set, Yaseen 2018~\cite{yaseen2018classification} (Y-18) is passed through a codec simulation at high compression to introduce distortions in the data to produce the Augmented Y-18 dataset.}
  \label{fig:Data Augmentation Strategy}
  \vspace{-10pt}
\end{figure}
Despite recent advances, heart sound classification research has been held back by the limited amount of clean, annotated heart murmur PCG data available to the public~\cite{e23060667}. Although there have been some attempts to address this in recent years, with new murmur datasets being made public~\cite{circor} and repositories like PhysioNet~\cite{PhysioNet} that hosts this data, the amount of PCG data available pales in comparison to other audio datasets such as AudioSet~\cite{audioset} and speech-specific ones like VoxCeleb~\cite{voxceleb}, where impressive performance has been achieved.

One of the ways the problem of limited data can be overcome is through data augmentation. Our data augmentation approach is outlined in Figure~\ref{fig:Data Augmentation Strategy} and the details of the implementation of the data augmentation is outlined in Seciton~\ref{subsec:codec}.

This work focuses on improving the performance of time-domain classifiers on the Yaseen 2018 (Y-18) dataset~\cite{yaseen2018classification} which is popular due to its quality and balanced representation of various murmurs. Since the dataset was published, the classification error rate (CER) of models on the Y-18 dataset under a 10-fold cross-validation (CV) approach has reached as low as 0.10 in the frequency domain case. However, the best model under the time-domain approach remains at 0.68 CER.

In this paper, we report the heart sound classification using the model M5~\cite{M5} through a time-domain approach which achieves a CER of 0.8 without data augmentation. Then we use the codec simulation data augmentation approach reported in~\cite{codec} and see an improvement in performance to a CER of 0.2. This outperforms the existing baseline of 0.68 and validates the use of codec simulation in augmenting PCG data.

\section{Methodology}

\subsection{Yaseen Dataset}
The Yaseen Dataset~\cite{yaseen2018classification} is a public dataset that consists of 1000 recordings of heart sounds evenly distributed across 5 categories, as shown in table~\ref{tab:Y18_data_distribution}. The 5 categories are: normal (N) aortic stenosis (AS) mitral stenosis (MS) mitral regurgitation (MR) and mitral valve prolapse (MVP). The data was collected by the authors of~\cite{yaseen2018classification} from myriad online sources and processed aligned through downsampling to 8kHz and conversion to single channel. Some of these sources include medical textbooks and online websites. The length of the audio files ranges from 1 second to 4 seconds. 

\begin{table}[h!]
  \caption{Distribution of the data in Y-18 across different categories}
  \label{tab:Y18_data_distribution}
  \centering
  \begin{tabular}{llc}    
    \textbf{Type} & \textbf{Category} & \textbf{Number of Samples} \\
    \hline \\
    Normal & Normal (N) & 200\\[0.5pt]
    Abnormal & Aortic Stenosis (AS) & 200\\[0.5pt]
    Abnormal & Mitral Stenosis (MS) & 200\\[0.5pt]
    Abnormal & Mitral Regurgitation (MR) & 200\\[0.5pt]
    Abnormal & Mitral Valve Prolapse (MVP) & 200\\[0.5pt]
  \end{tabular}
\end{table}

\begin{figure}[t]
  \centering
  \includegraphics[width=\linewidth]{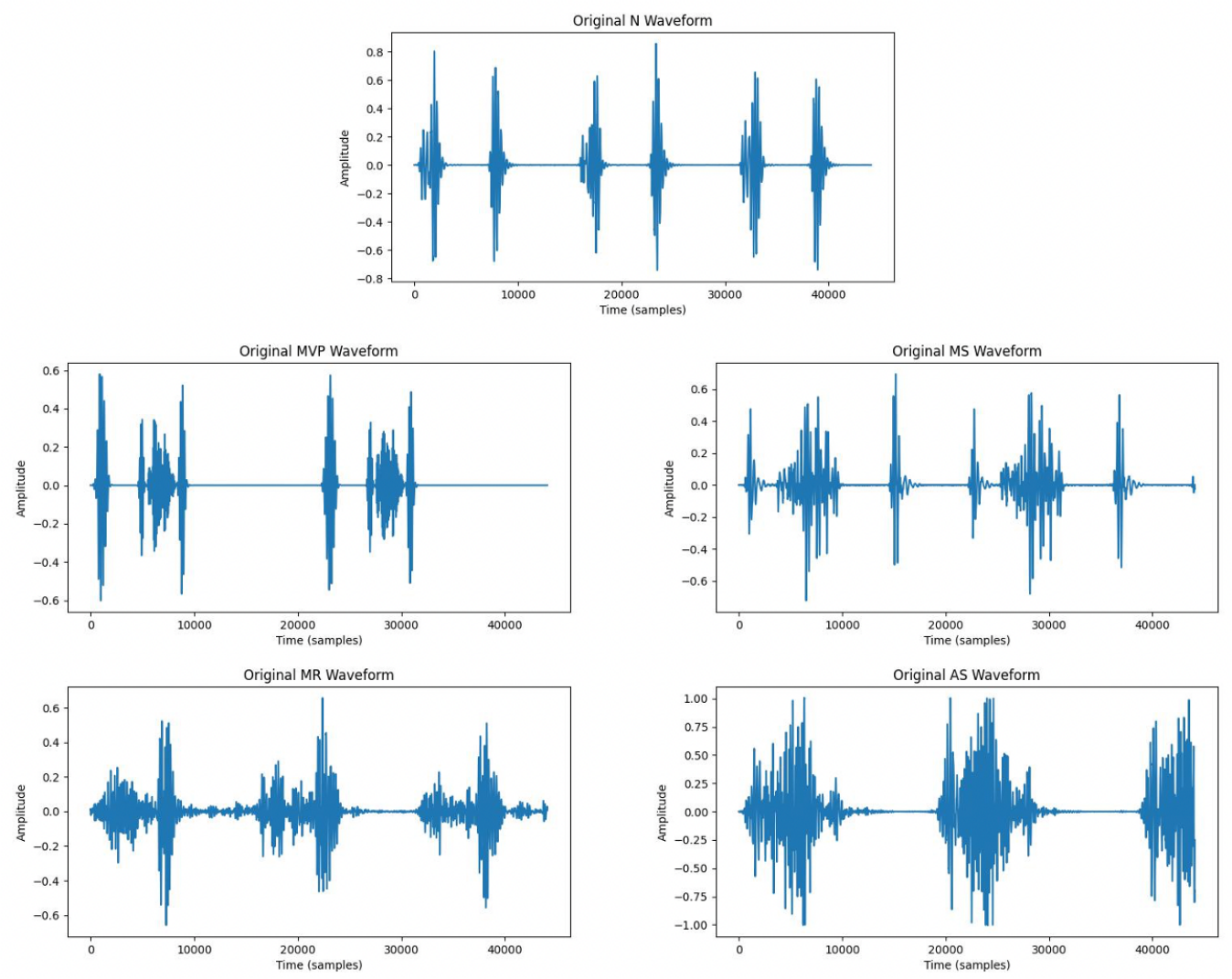}
  \caption{Waveform plots of the 5 categories of heart sounds. Each of the 5 categories, Normal (Top Row), Mitral Valve Prolapse (Middle Row Left), Mitral Stenosis (Middle Row Right), Mitral Regurgitation (Bottom Row Left), and Aortic Stenosis (Bottom Row Right) have features that distinguish them from each other. From the waveforms, we can also see that in the normal heart sound the S1 and S2 sounds are clearly visible, but for the abnormal heart sounds these S1 and S2 peaks cannot always easily be visually identified, especially in the case of AS and MR shown in the bottom row.}
  \label{fig:HSWave}
\end{figure}

Compared to other public datasets such as the PASCAL 2011~\cite{pascal-chsc-2011} and CirCor Digiscope 2022~\cite{circor} datasets, the Y-18 dataset offers the advantage of being a balanced dataset across each of the categories of heart murmurs. The different categories and the differences in their waveforms are shown in Figure~\ref{fig:HSWave}). Each of the categories has a distinct waveform which can be seen in the plot. A heartbeat consists of two peaks in the audio waveform, forming a "lub" and "dub" sound. These are referred to as the S1 and S2 peaks respectively. The S1 and S2 peaks can be clearly seen in the plot of the normal recording, while they are harder to spot in the abnormal cases. 

\begin{figure}[htbp!]
  \centering
  \includegraphics[width=\linewidth]{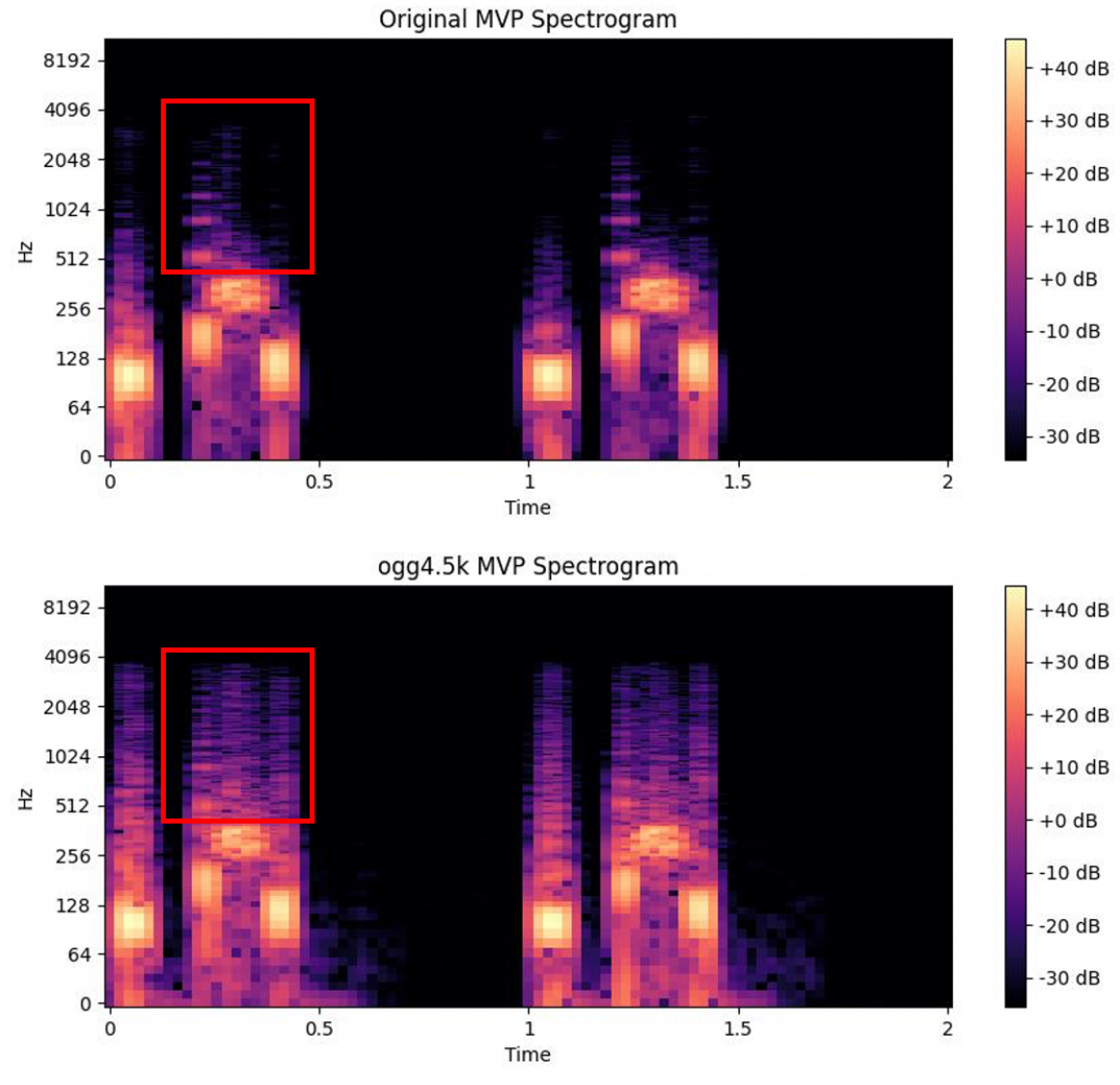}
  \caption{Comparison of a sample MVP PCG signal before and after the codec data augmentation. The most compressed ogg 4.5k bitrate codec is used here for illustration and the spectrogram is plotted for an easier visualization of the differences. The original spectrogram is shown in the top image while the spectrogram after the codec simulation is shown in the bottom image. After passing through the code there is more smearing in the spectrogram observed throughout the spectrogram. However, it is most obvious in the area highlighted in the red box, where the initial banding pattern can almost no longer be seen due to the increase in the noise.}
  \label{fig:dataAugCompare}
\end{figure}

\subsection{Codec data augmentation}
\label{subsec:codec}
The use of codec simulation to improve audio classification accuracy was first reported on an automatic speech recognition task by the authors of~\cite{codec}. It was found that by running an audio recording through a codec simulation, Word-Error-Rate (WER) could be improved by 7.28\% - 12.78\% when compared to a strong baseline baseline~\cite{codec}. 

In this work, codec augmentation is performed using the ffmpeg package to simulate the codec. The settings for the codec simulation used is the Opus (OGG) format with bitrates of 4.5k, 5.5k, and 7.7k. We make use of high compression codec with low bitrate to increase the level of distortion in the training data so that we can improve the overall classification accuracy. The codec simulation is implemented in a two-step process in the command line as follows:
\begin{verbatim}
ffmpeg -i <input_file>.wav -c:a libopus 
        -b:a <bitrate> temp_file.ogg
ffmpeg -i temp_file.ogg -ar 
        8000 <output_file>.wav
\end{verbatim}
The process can also be performed within a Python script by using the subprocess package. In this case, we used a Python script to loop over the list of chosen bitrates for all files in the Y-18 dataset to create the final augmented Y-18 dataset. While we only used the ogg codec in this study, this process can also be generalized to include more codecs. 

The distortion created by the codec simulation can be visualized using a spectrogram. In Figure~\ref{fig:dataAugCompare} we show the spectrogram of an MVP PCG signal in its original form compared to its distorted form. We can see that the codec simulation does indeed result in some smearing on the spectrogram and the loss of some of the PCG signals. This makes the task of the classifier more difficult and thus should guide it toward extracting more general features that are not impacted by the distortions. 

Overall, using the codec simulation at 3 different bitrates, we create 3 additional copies of the Y-18 training dataset resulting in 1000 original PCG recordings and 3000 augmented PCG recordings. All 4000 PCG recordings are used in the training of the model under the augmentation training regime. This set of data augmentations were selected based on the hardest settings reported by the authors of~\cite{codec} which we believe serves as the strongest augmentation.

\begin{figure*}[h!]
    \centering
    \includegraphics[width=420pt]{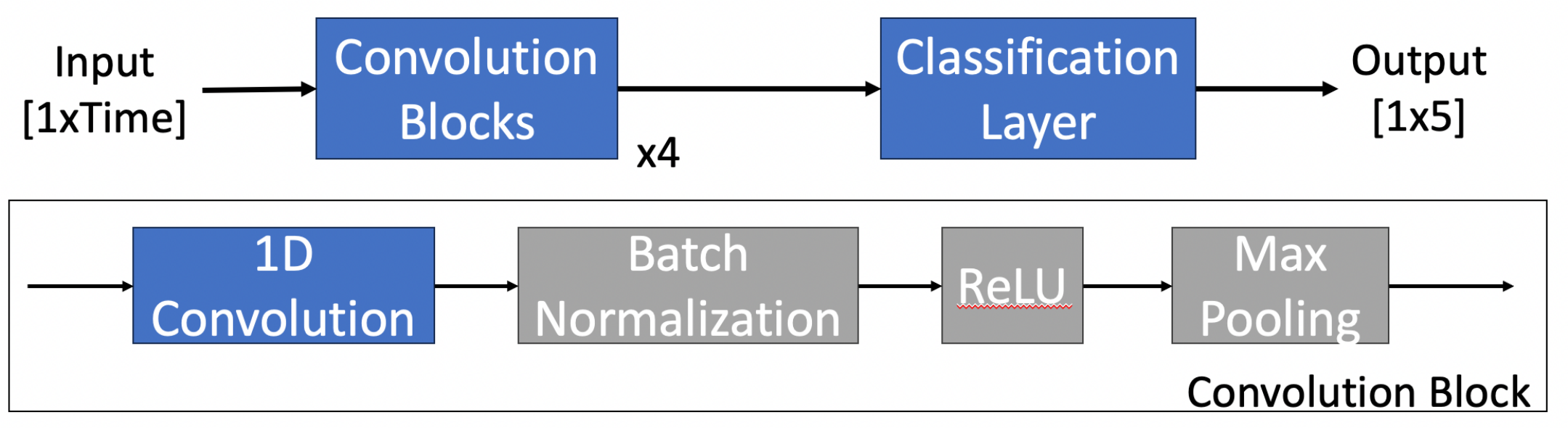} %420pt width is not bad
    \caption{Detailed view of the model used in the work. The M5 model consists of 4 convolution blocks followed by the classification later. Each convolution block consists of a single 1D convolutional layer, followed by batch normalization, a Rectified Linear Unit activation (ReLU), and max pooling. The parts of the model that have trainable parameters are indicated in blue with the non-trainable functions are indicated in grey.}
    \label{fig:model-diagram}
    % \vspace{-10pt}
\end{figure*}

\subsection{Model}
The model used in this work is a simple time-domain convolutional model. The model consists of 4 convolutional blocks, followed by a linear classifier label as shown in Figure~\ref{fig:model-diagram}. This architecture was first reported in~\cite{M5} but we have adapted the number of channels and the output size of the model to be suitable for the heart sound classification task. 

The output layers was set to 5 to match the number of classes in the dataset. The stride of the first convolutional layer was set to 16 with a kernel size of 80 to reduce the size of the input into the rest of the model and act as a time-domain encoder. The channel dimension of the first layer was set to 32 which is increased to 64 in the third layer. The pooling and batch normalization layers are implemented as per~\cite{M5}. Despite the simplicity of this model, it achieves impressive performance even when compared to a previously reported CNN-BiLSTM~\cite{JAMIL2023106734} model.

\subsection{Training Methodology}
For all results, we perform 10-fold cross-validation and report the average CER across all 10 folds, in alignment with the relevant comparison models in the literature. The models are trained on softmax and cross-entropy loss. Additionally, we downsample the recordings to 2kHz as a pre-processing step before passing the PCG signal to the model. The batch size used in the training is 5 and the optimizer used is the Adam optimizer~\cite{kingma2017adam} with a learning rate of 0.0005 and a weight decay of 0.0001.

On the use of data, this study compares the performance of two different training regimes: one without data augmentation and one with data augmentation. Firstly, in the unaugmented training regime, we use only the 1000 original PCG recordings from the Y-18 dataset, while in the augmented training regime, as mentioned in section~\ref{subsec:codec} we used a combination of the original Y-18 dataset and the Y-18 dataset after running through the codec at 3 different bitrates. 

During testing, the classification results on the original data and the data that has been run through the codec are calculated separately, to ensure compatibility with the results that have already been reported in the literature.

\section{Experiments}
\vspace{5pt}
\subsection{Comparison with baseline}
In this section, we compare the results of the M5 models under the two training configurations with the results reported in the literature. These comparisons are reported in Table~\ref{tab:comparison}.

The most common data sampling approach is cross-validation, with the original Y-18 authors~\cite{yaseen2018classification} using 5-fold cross-validation for their initial baselines, although subsequent authors have used a 10-fold cross-validation approach. Aside from cross-validation, the authors of \cite{nguyen2023heart} use a simple split of 70\% training and 30\% validation data split (70-30), which is probably not advisable for a small dataset like Y18. In contrast, \cite{JAMIL2023106734} uses an innovative multi-round transfer learning (MRTL) approach and performs numerous comparisons across multiple computer vision models and achieves very good results across all of them.
\begin{table}[h!]
  \caption{Comparison with Baselines on the Y18 Dataset}
  \label{tab:comparison}
  \centering
  \begin{tabular}{llcc}    
    \textbf{Model} & \textbf{Domain} & \textbf{Sampling Method} & \textbf{CER	$\downarrow$} \\
    \hline\hline \\
    MFCC+DWT KNN~\cite{yaseen2018classification} & Frequency & 5-fold CV & 2.60\\[0.5pt]
    MFCC+DWT SVM~\cite{yaseen2018classification} & Frequency & 5-fold CV & 2.10\\[0.5pt]
    MFCC+DWT DNN~\cite{yaseen2018classification} & Frequency & 5-fold CV & 7.9\\[0.5pt]
    LSTM~\cite{nguyen2023heart} & Frequency & 70-30 & 5.00\\[0.5pt]
    CNN~\cite{nguyen2023heart} & Frequency & 70-30 & 0.33\\[0.5pt]
    Multiple Models~\cite{JAMIL2023106734} & Frequency & MRTL & $<$ 2.00\\[0.5pt]
    \hline
    WaveNet~\cite{OH2020105604} & Time & 10-Fold CV & 3.00\\[0.5pt]
    BiLSTM~\cite{ALKHODARI2021105940} & Time & 10-Fold CV & 7.36\\[0.5pt]
    CNN~\cite{ALKHODARI2021105940} & Time & 10-Fold CV & 1.68\\[0.5pt]
    CNN-BiLSTM~\cite{ALKHODARI2021105940} & Time & 10-Fold CV & \textbf{0.68}\\[0.5pt]
    CardioXNet~\cite{cardioxnet} & Frequency & 10-Fold CV & 0.40\\[0.5pt]
    ViT~\cite{JAMIL2023106734} & Frequency & 10-Fold CV & \textbf{0.10}\\[0.5pt]
    \hline
    M5 & Time & 10-Fold CV & 0.80\\[0.5pt]
    M5 + Augmentation & Time & 10-Fold CV & \textbf{0.20}\\[0.5pt]
  \end{tabular}
\end{table}

Among the models that make use of 10-fold comparisons, where the results are comparable, there are two types of approaches. The time-domain approach uses the raw audio waveform for the classification. The frequency-domain approach first converts the audio waveform into a series of spectrograms, which can be a beneficial feature engineering step to improve the model performance. However, the frequency domain approach has a small disadvantage during implementation as the Fast Fourier Transform operation can sometimes be costly depending on the approach. Nevertheless, the frequency-domain approach currently outperforms the time-domain approach. The best time-domain approach using a CNN-BiLSTM model~\cite{ALKHODARI2021105940} has a CER of 0.68 while the best frequency-domain approach has a CER of only 0.1 using a Vision Transformer~\cite{JAMIL2023106734}.

The M5 model using the baseline training configuration underperforms the CNN-BiLSTM model with a CER of 0.80, however, with the codec simulation augmented training dataset the M5 model can outperform the CNN-BiLSTM model. This result thus shows the importance of data augmentation and the effectiveness of our codec simulation data augmentation approach.

\subsection{Analysis Across Codec and Original testing datasets}
In this section, we report the performance of the M5 model under both training configurations and their respective validation sets. While training is performed with both the original and augmented data, the testing is done on the original data and the codec data separately to maintain comparability with the previous models. The results of these experiments are shown in Table~\ref{tab:originalvscodec}.

\begin{table}[h!]
  \caption{Performance on the Original and Codec datasets}
  \label{tab:originalvscodec}
  \centering
  \begin{tabular}{lcc}    
    \textbf{Model} & \textbf{Original CER	$\downarrow$} & \textbf{Codec CER	$\downarrow$} \\
    \hline\hline \\
    M5  & 0.80 & 1.63 \\[0.5pt]
    M5 + Augmentation & \textbf{0.20} & \textbf{0.57}\\[0.5pt]
  \end{tabular}
\end{table}

In the training configuration with the M5 model and no data augmentation, we obtain an original CER of 0.8 and codec CER of 1.63. In this case, the model has seen the original Y-18 data but not the codec augmented data. The performance difference between these two CERs is likely due to the distortions introduced by the codec simulation.

In the training configuration with the M5 model and data augmentation, we see that both the original CER and the codec CER improves. On the codec CER, the large performance improvement from 1.63 to 0.57 is likely due to the model having now seen the codec data during its training as well. On the other hand, the improvement of the original CER shows that the codec-augmented data in the training of the model can help guide the model towards using better and more general features in the classification, which improves its generalization performance.

\section{Discussion}
\vspace{5pt}
The result of the M5 model on the Y-18 dataset as reported in Table~\ref{tab:comparison} shows that a simple deep convolutional neural network using the time-domain approach can be competitive with much more complicated models like the Vision Transformer. This also brings the time-domain approach to a level that is competitive with the frequency-domain approach like Vision Transformer. In future work, we intend to attempt classification using transformer-based time-domain approaches such as ACA-Net~\cite{yip2023acanet}. 

The high performance of all models across the literature suggests that there is room for further increasing the dataset size to make the task more difficult. One way this can be done is through training the model on the Y-18 dataset and testing the model on other heart sound datasets that have been collected. This however creates some significant issues due to out-of-domain noise profiles, but can be an avenue for further research as well.

Furthermore, it would be beneficial to evaluate the models on real-world, clinical data to assess their performance in practical settings. Clinical data often presents additional challenges such as varying recording conditions, patient demographics, and the presence of other pathological conditions. Evaluating the models under these conditions would provide a more realistic assessment of their effectiveness.

\section{Conclusion}
\vspace{5pt}
In this work we have shown that data augmentation of heart sounds through codec simulation is an effective method for improving the classification of heart sounds on the Y-18 dataset. Using the M5 model, we also show that it is possible to improve the accuracy of the time-domain classification approach to be competitive with the frequency-domain models. Spcifically, our data augmentation strategy improves the CER of the M5 model from 0.8 to 0.2. On transmitted audio segments the improvement is even greater from 1.63 to 0.57. Overall this validates the codec simulation approach as an effective data augmentation approach towards addressing the problem of limited data availability in the field of heart sound classification.

\section{Acknowledgements}
This research is supported by ST Engineering Mission Software \& Services Pte. Ltd under a collaboration programme (Research Collaboration No: REQ0149132). We would like to acknowledge the High Performance Computing Centre of Nanyang Technological University Singapore, for providing the computing resources, facilities, and services that have contributed significantly to this work.

\bibliographystyle{IEEEtran}
\bibliography{conference_101719}

\end{document}